\documentclass[12pt]{article}
\usepackage[cmtip,arrow]{xy}\usepackage{pb-diagram,pb-xy}%

\input xy
\xyoption{all}
\input xy
\xyoption{all}
\usepackage{heck}
\usepackage{cite}
\usepackage{graphicx}
\usepackage{makeidx}
\usepackage{multicol}
\usepackage{amsfonts}
\usepackage{mathrsfs}
\usepackage{amssymb}
\usepackage{amsmath}%

\providecommand{\U}[1]{\protect\rule{.1in}{.1in}}
\numberwithin{equation}{section}

\newcommand{\bea}{\begin{eqnarray}}
\newcommand{\eea}{\end{eqnarray}}
\newcommand{\be}{\begin{equation}}
\newcommand{\ee}{\end{equation}}

\newcommand{\bem}{\begin{pmatrix}}
\newcommand{\eem}{\end{pmatrix}}

\def\U{\Upsilon}

\def\cn{{\cal N}}

\def\cp{{\cal P}}

\def\cs{{\cal S}}
\def\ct{{\cal T}}

\def\cw{{\cal W}}

\def \Z {{\mathbb Z}}
\def \C {{\mathbb C}}

\xyoption{arc}

\date{October, 2012}

\institution{SISSA}{\centerline{SISSA, via Bonomea 265, I-34100 Trieste, ITALY}}

\title{Infinitely many $\cn=2$ SCFT\\
 with $ADE$ flavor symmetry}
\authors{Sergio Cecotti\footnote{e-mail: {\tt cecotti@sissa.it}} and Michele Del Zotto\footnote{e-mail: {\tt eledelz@gmail.com}}%
}

\abstract{We present evidence that for each $ADE$ Lie group $G$ there is an infinite tower of 4D $\cn=2$ SCFTs, which we label as $D(G,s)$ with $s\in\mathbb{N}$, having (at least) flavor symmetry $G$. For $G=SU(2)$, $D(SU(2),s)$ coincides with the Argyres--Douglas model of type $D_{s+1}$, while for larger flavor groups the models are new (but for a few previously known examples). 
When its flavor symmetry $G$ is gauged, $D(G,s)$ contributes to the Yang--Mills beta--function as $\tfrac{s}{2(s+1)}$ adjoint hypermultiplets.

The argument is based on a combination of Type IIB geometric engineering and the categorical deconstruction of {\ttfamily arXiv:1203.6743}. One first engineers a class of $\cn=2$ models which, trough the analysis of their category of quiver representations, are identified as asymptotically--free gauge theories with gauge group $G$ coupled to some conformal matter system. Taking the limit $g_\mathrm{YM}\rightarrow 0$ one isolates the matter SCFT which is our $D(G,s)$.   
}

\begin{document}

\maketitle


\section{Introduction}\label{intt}

One of the most remarkable aspects of extended supersymmetry is the possibility of constructing and studying in detail many four--dimensional SCFTs which do not have any (weakly coupled) Lagrangian formulation and hence are intrinsically strongly coupled. The prototype of such theories are given by the Argyres--Douglas $\cn=2$ models \cite{AD}, which have an $ADE$ classification; those of type $D_r$ ($r=2,3,\cdots$) have a $SU(2)$ global symmetry which may be gauged \cite{CV11}. Other important classes of $\cn=2$ SCFTs are the so--called class--$\cs$ theories \cite{Gaiotto,GMN09}, the $(G,G^\prime)$ models of \cite{CNV} ($G,G^\prime$ being a pair of $ADE$ groups), and their generalizations \cite{arnold1,arnold2,arnold3}.

Of particular interest are the $\cn=2$ SCFT with an exceptional flavor symmetry, $E_6,E_7,E_8$. Here the basic examples are the Minahan--Nemeschansky (MN) models \cite{MN1,MN2} (see also \cite{seii,ben}); the flavor symmetry alone rules out any weakly coupled description; for instance, if we gauge the $E_8$ symmetry of the last MN model we get a contribution to the $\beta$--function which is $1/10$ of an hypermultiplet in the minimal representation (the adjoint) \cite{E8beta}.

The purpose of this letter is to present evidence for the existence of infinitely many such SCFT. For each $ADE$ Lie group $G$ --- in particular, for $E_6$, $E_7$, and $E_8$ --- we have an infinite tower of models with (at least) $G$ flavor symmetry. For a given $G$, the models are labelled by a positive integer $s\in\mathbb{N}$. We denote these models as $D(G,s)$.
 When coupled to $G$ SYM, $D(G,s)$ will contribute to the YM $\beta$--function as $$\frac{s}{2(s+1)}\times \text{(adjoint hypermultiplet).}$$  
This implies that $D(G,s)$ cannot have a Lagrangian formulation except for sporadic, very special, pairs $(G,s)$. While these sporadic Lagrangian models are not new theories, they are quite useful for our analysis because, in these special cases, we may check our general results against standard weak coupling computations, getting perfect agreement.

In simple terms our construction is based on the following ideas (see ref.\!\!\cite{CNV} for the general set--up). We start by considering the `compactification' of Type IIB on the local Calabi--Yau hypersurface of equation
\begin{equation}\label{kkkam45}
W_{G,s}(z,x_1,x_2,x_3)=\Lambda^b\,e^{(s+1)z}+\Lambda^b\,e^{-z}+W_G(x_1,x_2,x_3),
\end{equation}
where $W_G(x_1,x_2,x_3)$ stands for (the versal deformation of the) minimal $ADE$ singularity of type $G$. Seen as a 2d superpotential, $W_{G,s}$ corresponds to a model with central charge $\hat c$ at the UV fixed point equal to 
$\hat c_\mathrm{uv}= 1+\hat c_G <2,$,
where $\hat c_G$ is the central charge of the minimal $(2,2)$ SCFT of type $G$. Since $\hat c_\mathrm{uv}<2$, the criterion of the $2d/4d$ correspondence \cite{CNV,GVW} is satisfied, and we get a well--defined QFT in 4D. 
For $s=0$ the theory we get is just pure SYM with gauge group $G$ \cite{CNV}. By the usual argument (see \textit{e.g.} \cite{tack, CNV,CV11}) for all $s\in\mathbb{N}$ the resulting 4D theory is UV asymptotically free; in facts, it is SYM with gauge group $G$ coupled to some matter which is `nice' in the sense of \cite{tack}, that is, it contributes to the YM $\beta$--function less than half an adjoint hypermultiplet. 

Taking the limit $g_\mathrm{YM}\rightarrow 0$, we decouple the SYM sector and isolate the matter theory that we call $D(G,s)$. It is easy to see that this theory should be conformal. Indeed, the `superpotential' \eqref{kkkam45} is the sum of two decoupled terms; at the level of the BPS quiver of the 4D $\cn=2$ theory, this produces the \emph{triangle tensor product} \cite{kellerP} of the quivers $\widehat{A}(s+1,1)$ and $G$ (compare, for $s=0$, with the pure SYM case \cite{CNV,cattoy}).  The decoupling limit affects
only the first factor in the triangle product, so, roughly speaking, we expect 
\begin{equation}\label{ttnnhh}D(G,s)\equiv \text{(something depending only on }s)\boxtimes G.\end{equation}
Modulo some technicality, this is essentially correct. Then, from the $2d/4d$ correspondence, it is obvious that the resulting theory is UV conformal iff `(something depending only on $s)$'
is. This can be settled by setting $G=SU(2)$. In this case $D(SU(2),s)$ is Argyres--Douglas of type $D_{s+1}$ \cite{CV11,cattoy} which is certainly UV superconformal. Hence $D(G,s)$ is expected to be superconformal for all $G$ and $s$. (Below we shall be more specific about the first factor in the \textsc{rhs} of \eqref{ttnnhh}.)
Alternatively, we can argue as follows: the gauge theory engineered by the CY hypersuface \eqref{kkkam45} has just one essential scale, $\Lambda$; the decoupling limit $g_\mathrm{YM}\rightarrow 0$ corresponds to a suitably defined scaling limit $\Lambda\rightarrow 0$; therefore we should end up to the UV--fixed point SCFT.

The construction may in principle be extended by considering the triangle tensor products of two affine theories, $\widehat{H}\boxtimes\widehat{G}$, which are expected to be asymptotically--free $\cn=2$ theories with non--simple gauge groups. 

Technically, the analysis of the decoupling limit is based on the `categorical' classification program of  4D $\cn=2$ theories advocated in ref.\!\!\cite{cattoy}. In the language of that paper, our problem is to construct and classify the non--homogeneous $G$--tubes by isolating them inside the light subcategory of the 4D gauge theory.
\medskip

The rest of this letter is organized as follows. In section 2 we briefly review some material we need. In section 3 we analyze the 4D gauge theories of the form $\widehat{H}\boxtimes G$: we study both the strong coupling and the weak coupling. We also discuss some examples in detail. In section 4 we decouple the SYM sector and, isolate the $D(G,s)$ SCFT, and describe some of their physical properties. In section 5 we sketch the extensions to  the $\widehat{H}\boxtimes\widehat{G}$ models.
 Technical details and more examples are confined in the appendices. 

\section{Brief review of some useful facts}

We review some known facts we need. Experts may prefer to jump to section 3.
For the basics of the quiver representation approach to the BPS spectra of 4D $\cn=2$ theories we refer to\cite{CV11,ACCERV1,ACCERV2}\!\!\cite{cattoy}.

\subsection{AF $\cn=2$ $SU(2)$ gauge theories and Euclidean algebras}\label{AFEucli}

We shall be sketchy, full details may be found in \cite{CV11} and \cite{cattoy}.

The full classification of the $\cn=2$ $SU(2)$ gauge theories whose gauge group is strictly $SU(2)$ and which are both complete and asymptotically--free is presented in ref.\!\!\cite{CV11}. Such theories are in one--to--one correspondence with the mutation--classes of quivers obtained by choosing an acyclic orientation of an affine $\widehat{A}\widehat{D}\widehat{E}$ Dynkin graph. For $\widehat{D}_r$ ($r\geq 4$) and $\widehat{E}_r$ ($r=6,7,8$) all orientations are mutation equivalent, while in the $\widehat{A}_r$ case the inequivalent orientations are characterized by the net number $p$ (resp.\!\! $q$) of arrows pointing in the clockwise (anticlockwise) direction along the cycle; we write $\widehat{A}(p,q)$ for the $\widehat{A}_{p+q-1}$ Dynkin graph with such an orientation ($p\geq q\geq 1$). The case $\widehat{A}(p,0)$ is different because there is a closed \textit{oriented} $p$--loop. The corresponding path algebra $\C\widehat{A}(p,0)$ is infinite--dimensional, and it must be bounded by some relations which, in the physical context, must arise from the gradient of a superpotential, $\partial\cw=0$ \cite{ACCERV1,ACCERV2}. For generic $\cw$,  $\widehat{A}(p,0)$ is mutation--equivalent to the $D_p$ Argyres--Douglas model \cite{CV11,cattoy} which has an $SU(2)$ global symmetry. By the triality property of $SO(8)$, the $D_4$ Argyres--Douglas model is very special: its flavor symmetry gets enhanced to $SU(3)$ --- this exception will be relevant below. 

One shows \cite{CV11,cattoy} that these $\cn=2$ affine theories correspond to $SU(2)$ SYM gauging the global $SU(2)$ symmetries of a set of Argyres--Douglas models of type $D_r$ as in the table
\begin{equation}\label{tabletable}
\begin{tabular}{|c|c|}\hline
acyclic affine quiver $\widehat{H}$ & matter content\\\hline
$\widehat{A}(p,q)$\quad $p\geq q\geq1$ & $D_p\oplus D_q\ (\oplus D_1)$\\\hline
$\widehat{D}_r$\quad $r\geq 4$& $D_2\oplus D_2\oplus D_{r-2}$\\\hline
$\widehat{E}_r$\quad $r=6,7,8$& $D_2\oplus D_3 \oplus D_{r-3}$\\\hline 
\end{tabular}
\end{equation}
where $D_1$ stands for the empty matter and $D_2\equiv A_1\oplus A_1$ for a free hypermultiplet doublet. The Type IIB geometry which engineers the $\cn=2$ model associated to each acyclic affine quiver in the first column is described in ref.\!\!\cite{CV11}. For instance, for $\widehat{A}(p,q)$ the geometry is
\begin{equation}\label{xxxaaaq}
W_{p,q}(z,x_i)\equiv \Lambda^b\,e^{p z}+\Lambda^b\,e^{-q z}+x_1^2+x_2^2+x_3^2=0.
\end{equation}
One also shows \cite{CV11,cattoy} that the contribution of each $D_r$ matter system to the $SU(2)$ YM $\beta$--function coefficient $b$
$$\mu\frac{\partial}{\partial\mu}\frac{4\pi}{g^2_\mathrm{YM}}=\frac{b}{2\pi},$$
is given by
\begin{equation}
b(D_r)=\frac{2(r-1)}{r}
\end{equation}
Using this formula, one checks \cite{CV11} that the models listed in \eqref{tabletable} precisely correspond to all possible (complete) matter systems which are compatible with asymptotic freedom. 

For our purposes it is important to describe the decoupling process of the matter from the SYM sector; it is described in terms of the BPS spectrum in ref.\!\!\cite{cattoy}. The BPS states which have a bounded masses in the limit $g_\mathrm{YM}\rightarrow 0$ are precisely the BPS particles with zero magnetic charge. In terms of the representations of the acyclic affine quiver $\widehat{H}$ these light states correspond to the ones having vanishing Dlab--Ringel defect \cite{RI,CB}. 
To describe the BPS states which remain light in the decoupling limit, one introduces the Abelian (sub)category of the \textit{light} representations\footnote{\ See also \S.\,\ref{sec:light} below.} \cite{cattoy}, which --- in the affine case --- precisely corresponds to the category of the regular representation \cite{RI,CB}. This category has the form \cite{RI}
\begin{equation}\label{ringel}
\ct=\bigvee_{\lambda\in\mathbb{P}^1}\ct_\lambda,
\end{equation}
where the $\ct_\lambda$ are stable periodic tubes; for generic $\lambda$, $\ct_\lambda$ is a homogeneous tube ($\equiv$ period 1) \cite{RI}. This, in particular, means that for these affine models the light BPS states consists of a single vector--multiplet, the $W$ boson, plus finitely many hypermultiplets, which are the BPS states of the matter system (the matter spectrum at $g_\mathrm{YM}\sim 0$ depends on the particular BPS chamber). It follows that the matter sector corresponds to the \emph{rigid bricks}\footnote{\ A representation $X$ is a \emph{brick} iff $\mathrm{End}\,X=\C$, and it is \emph{rigid} if, in addition, $\mathrm{Ext}^1(X,X)=0$.} of $\ct$ \cite{cattoy}. The rigid bricks belong to the finitely--many tubes $\ct_\lambda$ which are not homogeneous. It is well--known that for each affine quiver there is precisely one non--homogeneous tube of period $p_i$ for each $D_{p_i}$ matter subsystem in the second column of table  \eqref{tabletable}.
To show that the matter isolated by the decoupling process is the combination of Argyres--Douglas models in table \eqref{tabletable} one may use either rigorous mathematical methods or physical arguments. Let us recall the mathematical proof \cite{cattoy}. The quiver of the matter category associated to a tube of period $p$ is obtained by associating a node $\bullet_i$ to each simple representation $S_i$ in the tube and connecting two nodes $\bullet_i$, $\bullet_j$ by $\mathrm{Ext}^1(S_i,S_j)$ arrows. The $\mathrm{Ext}^1(S_i,S_j)$ is easily computed with the help  of the $\Z_p$ symmetry of the periodic tube; the resulting quiver is then a single oriented cycle of length $p$. 
The same results may be obtained on physical grounds as follows 
(say for the case $\widehat{A}(p,q)$): in eqn.\eqref{xxxaaaq},
 $\Lambda$ stands for the scale set by asymptotic freedom, as specified by the asymptotic behavior  of the complex YM coupling \cite{tack}
\begin{equation}
\tau(a)\approx \frac{b}{2\pi i}\,\log\frac{\Lambda}{a}.
\end{equation}
The limit $g_\mathrm{YM}\rightarrow 0$ is $\Lambda\rightarrow 0$. 
We may take this limit keeping fixed either $pz+b\log \Lambda$ or $qz-b\log\Lambda$. These two limits correspond, respectively, to considering the local geometry of the hypersurface \eqref{xxxaaaq} around $z\sim+\infty$ and $z\sim -0$, which are precisely the two poles of the $\mathbb{P}^1$ with affine coordinate $e^z$; this $\mathbb{P}^1$ is identified  with the index set in eqn.\eqref{ringel} (and also with the Gaiotto plumbing cylinder \cite{cattoy}). Now it is clear that as $g_\mathrm{YM}\rightarrow 0$ we get  two \textit{decoupled} physical systems described by the geometries
\begin{gather}\label{heuristic}
e^{p z^\prime} +\text{quadratic}=0,\qquad e^{q z^\prime}+\text{quadratic}=0,
\end{gather}
which (formally at least) correspond to $\widehat{A}(p,0)$ and $\widehat{A}(q,0)$, respectively. The periodicity $\mod p,q$ of the two periodic tubes then corresponds to $$\exp(z)\rightarrow e^{2\pi i/p}\exp(z)\quad \text{and}\quad \exp(z^\prime)\rightarrow e^{2\pi i/q}\exp(z^\prime).$$

The cyclic quiver $\widehat{A}(p,0)$ should be supplemented by a superpotential $\cw$. The correct $\cw$ is easy to compute \cite{cattoy}: $\cw$ is just the $p$--cycle itself. The pair $(\widehat{A}(p,0), \cw=p\text{--cycle})$ is mutation--equivalent to a $D_p$ Dynkin quiver \cite{CV11}, and hence the matter system consists of one $D_p$ Argyres--Douglas system per each (non--homogeneous) tube of period $p$ in the family \eqref{ringel}. This gives table \eqref{tabletable}.  

\subsection{Triangle tensor products of $\cn=2$ theories}\label{triangletr}

This subsection is based on \cite{CNV,kellerP} and \S.\,10.1 of \cite{cattoy}.
Suppose we set Type IIB on a local CY hypersurface of the form
\begin{equation}\label{supersiper}
W(x_i,y_j)\equiv W_1(x_i)+W_2(y_j)=0.\end{equation}
From the $2d/4d$ correspondence \cite{CNV}, we know that this geometry defines a good 4D $\cn=2$ QFT provided the $(2,2)$ LG model defined by the superpotential $W(x_i,y_j)$ has $\hat c<2$ at the UV fixed point. In this case the 4D BPS quiver has incidence matrix\footnote{\ The incidence matrix $B$ of a $2$--acyclic quiver $Q$ is defined by setting $B_{ij}$ equal to the number of arrows from node $i$ to node $j$, a negative number meaning arrows in the opposite direction $i\leftarrow j$. $B$ is then automatically skew--symmetric.}
\begin{equation}
B=S^t-S,
\end{equation}
where $S$ is the Stokes matrix encoding the BPS spectrum of the (2,2) LG model \cite{CV92}. For superpotentials of the special form
\eqref{supersiper} the 2d theory is the product of two totally decoupled LG models, and hence the BPS spectrum of the 2d theory may be obtained as a `product' of the ones for the decoupled models, $S=S_1\otimes S_2$. This gives the incidence matrix for $W$ 
$$B=S_1^t\otimes S_2^t-S_1\otimes S_2.$$
The corresponding operation at the level of quivers is called 
the \emph{triangle tensor product} \cite{CNV}.

It is convenient to give an algebraic interpretation of this `product' of (2,2) LG theories which fixes the associated superpotential $\cw$ \cite{kellerP,arnold1,cattoy}. 
We assume that the quivers $Q_1$ and $Q_2$ of the  (2,2) LG theories $W_1$, $W_2$ are \textit{acyclic} --- hence, by classification \cite{CV92,CV11}, either orientations of $ADE$ Dynkin graphs or acyclic orientations of $\widehat{A}\widehat{D}\widehat{E}$ affine graphs. Let $\C Q_1$, $\C Q_2$ be the corresponding path algebras. We can consider the tensor product algebra
$\C Q_1\otimes\C Q_2$ spanned, as a vector space, by 
the elements $\alpha\otimes\beta$ and endowed with the product
\begin{equation}\label{prodprod}
\alpha\otimes\beta \cdot \gamma\otimes\delta= \alpha\gamma\otimes \beta\delta.
\end{equation}
Let $e_i$, (resp.\! $e_a$) be the lazy paths ($\equiv$ minimal idempotents) of the algebra $\C Q_1$ (resp.\! $\C Q_2$). The minimal idempotents of the tensor product algebra are $e_{ia}=e_i\otimes e_a$; for each such idempotent $e_{ia}$ there is a node in the quiver of the algebra $\C Q_1\otimes\C Q_2$ which we denote by the same symbol. The arrows of the quiver are\footnote{\ Here $s(\cdot)$ and $t(\cdot)$ are the maps which associate to an arrow its source and target node, respectively.}
\begin{equation}
e_i\otimes\beta \colon e_i\otimes e_{s(\beta)}\rightarrow e_i\otimes e_{t(\beta)},\qquad \alpha\otimes e_a \colon e_{s(\alpha)}\otimes e_a\rightarrow e_{t(\alpha)}\otimes e_a.
\end{equation}
However, there are non--trivial relations between the paths; indeed the product \eqref{prodprod} implies the commutativity relations 
\begin{equation}\label{commmcom}
e_{t(\alpha)}\otimes \beta \cdot \alpha\otimes e_{s(\beta)}=\alpha\otimes e_{t(\beta)}\cdot e_{s(\alpha)}\otimes\beta.
\end{equation}
In the physical context all relations between paths should arise in the Jacobian form $\partial\cw=0$ from a superpotential. In order to set the commutativity relations in the Jacobian form, we have to complete our quiver by adding an extra arrow for each pairs of arrows $\alpha\in Q_1$, $\beta\in Q_2$
\begin{equation}
\psi_{\alpha,\beta}\colon e_{t(\alpha)}\otimes e_{t(\beta)}\rightarrow
e_{s(\alpha)}\otimes e_{s(\beta)},
\end{equation}
and introducing a term in the superpotential of the form
\begin{equation}\label{commtermms}
\cw=\sum_{\text{pairs }\alpha,\beta}\psi_{\alpha,\beta}\Big(e_{t(\alpha)}\otimes \beta \cdot \alpha\otimes e_{s(\beta)}-\alpha\otimes e_{t(\beta)}\cdot e_{s(\alpha)}\otimes\beta\Big)
\end{equation}
enforcing the commutativity conditions \eqref{commmcom}.
The resulting completed quiver, equipped with this superpotential, is called the \emph{triangle tensor product} of $Q_1$, $Q_2$, written $Q_1\boxtimes Q_2$ \cite{kellerP}\!\!\!\cite{arnold1,cattoy}. 
\medskip

\textbf{Examples.} If both $Q_1$, $Q_2$ are Dynkin quivers their tensor product corresponds to the $(G,G')$ models constructed and studied in \cite{CNV}. If $Q_1$ is the Kronecker (affine) quiver $\widehat{A}(1,1)$ and $Q_2$ is a Dynkin quiver of type $G$, $\widehat{A}(1,1)\boxtimes G$ is the quiver (with superpotential) of pure SYM with gauge group $G$ \cite{CNV,ACCERV2,cattoy}.
\medskip

Although mathematically the procedure starts with two acyclic quivers, formally we may repeat the construction for any pair of quivers, except that the last step, the determination of $\cw$, may be quite tricky. When one factor, say $Q_2$, is acyclic there is a natural candidate for the superpotential on the $Q_1\boxtimes Q_2$ quiver: $\cw_\mathrm{cand.}$ is the sum of one copy the superpotential of $Q_1$ per node of $Q_2$, plus the terms \eqref{commtermms} implementing the commutativity relations.

\subsection{The light subcategory $\mathscr{L}$ and $G$--tubes}\label{sec:light}

Suppose we have a $\cn=2$  theory, which is a quiver model in the sense of \cite{CV11,ACCERV1, ACCERV2} and behaves, in some duality frame, as SYM with gauge group $G$ coupled to some `matter' system. We fix a quiver $Q$ which `covers' the region in parameter space corresponding to weak $G$ gauge coupling. Then there is a set of one--parameter families of representations of the quiver $Q$, $X_i(\lambda)$, $i=1,2\dots, \mathrm{rank}\,G$, which correspond to the simple $W$--boson vector--multiplets of $G$. Let $\delta_i=\dim X_i(\lambda)$ be the corresponding charge vectors. The magnetic charges of a representations $X$ are then defined by \cite{cattoy,half,nonsimply} 
\begin{equation}\label{thhffnnc}
m_i(X)=-C^{-1}_{ij} \,\langle \delta_j,\dim X\rangle_\text{Dirac},
\end{equation}
where $C$ is the Cartan matrix of the gauge group $G$ and 
the skew--symmetric integral bilinear form $\langle\cdot,\cdot\rangle_\text{Dirac}$ is defined by the exchange matrix $B$ of the quiver $Q$.

States of non--zero magnetic charge have masses of order $O(1/g^2_\mathrm{YM})$ as $g_\mathrm{YM}\rightarrow 0$, and decouple in the limit. Thus the BPS states which are both stable and light in the decoupling limit must correspond to quiver representations $X$ satisfying the two conditions: 1) $m_i(X)=0$ for all $i$; 2) if $Y$ is a subrepresentation of $X$, then $m_i(Y)\leq 0$ for all $i$. The subcategory of all representations satisfying these two conditions is an exact closed Abelian subcategory $\mathscr{L}$ which we call the \emph{light} category of the theory (w.r.t.\! the chosen duality frame).

If the gauge group $G$ is \emph{simple} the light category has a structure similar to the one in eqn.\eqref{ringel}; indeed \cite{cattoy}
\begin{equation}\label{tttbbfda}
\mathscr{L}=\bigvee_{\lambda\in\mathbb{P}^1}\mathscr{L}_\lambda,
\end{equation}
where the Abelian categories $\mathscr{L}_\lambda$ are called $G$--tubes. Almost all $G$--tubes in eqn.\eqref{tttbbfda} are homogeneous, that is, isomorphic to the ones for pure SYM with group $G$.
The matter corresponds to the (finitely many) $G$--tubes in eqn.\eqref{tttbbfda} which are \emph{not} homogeneous. Just as in \S.\,\ref{AFEucli}, there is a finite set of points $\lambda_i\in\mathbb{P}^1$ such that the $G$--tube $\mathscr{L}_{\lambda_i}$ is not homogeneous, and we can limit ourselves to consider one such $G$--tube at the time, since  distinct $G$--tubes correspond at $g_\mathrm{YM}=0$ to decoupled matter sectors (\!\cite{cattoy} or apply the physical argument around eqn.\eqref{heuristic} to the hypersurface \eqref{kkkam45}).

A very useful property of the light category $\mathscr{L}$, proven in different contexts \cite{cattoy,half,nonsimply}, is the following. Assume our theory has, in addition to $g_\mathrm{YM}\rightarrow 0$, a decoupling limit (\textit{e.g.}\! large masses, extreme Higgs breaking), which is compatible with parametrically small YM coupling $g_\mathrm{YM}$, and such that the decoupled theory has support in a subquiver\footnote{\ As explained in \cite{half}, this happens whenever the controlling function of the corresponding subcategory \cite{cattoy} is non--negative on the positive cone in $K_0(\mathsf{mod}\C Q)$ of actual representations.}  $\widetilde{Q}$ of $Q$. Then
\begin{equation}\label{tttrrxzaq}
X\in\mathscr{L}(Q)\quad\Rightarrow\quad X\big|_{\widetilde{Q}}\in \mathscr{L}(\widetilde{Q}),
\end{equation} 
a relation which just expresses the compatibility of the decoupling
limit with $g_\mathrm{YM}\sim 0$. This fact is quite useful since it allows to construct recursively the category $\mathscr{L}$ for complicate large quivers from the light categories associated to smaller quivers.
The light category $\mathscr{L}$ has a quiver (with relations) of its own. However, while typically a full non--perturbative category has a $2$--acyclic quiver, the quiver of a light category has, in general, both loops and pairs of opposite arrows $\leftrightarrows$ (see examples in \cite{cattoy,half,nonsimply}). It depends on the particular superpotential $\cw$ whether the pairs of opposite arrows may or may not be integrated away.

\section{The $\cn=2$ models $\widehat{H}\boxtimes G$}

We consider the triangle tensor product $\widehat{H}\boxtimes G$ where $\widehat{H}$ stands for an acyclic affine quiver (listed in the first column of table \eqref{tabletable}), and $G$ is an $ADE$ Dynkin quiver. Since $\hat c(\widehat{H})=1$ and $\hat c(G)<1$, the total $\hat c$ is always less than 2, and thus all quivers of this form correspond to good $\cn=2$ QFT models.
If $\widehat{H}=\widehat{A}(1,1)$, the model $\widehat{H}\boxtimes G$ correspond to pure $\cn=2$ SYM with group $G$. In figure \ref{kkkam45} we show the  quiver (with superpotential)  corresponding to the simplest next model \textit{i.e.}\!  $\widehat{A}(2,1)\boxtimes A_2$, the general case being a repetition of this basic structure\footnote{\ For $\widehat{H}=\widehat{A}(p,p)$, $\widehat{D}_r$ and $\widehat{E}_r$ we have an equivalent square product quiver without `diagonal' arrows; for
$\widehat{A}(p,q)$ we may reduce to a quiver with just $p-q$ diagonal arrows.}. We call the full subquiver $\widehat{H}\boxtimes\{\bullet_a\}\subset \widehat{H}\boxtimes G$ `the affine quiver over the $a$--th node of the Dynkin graph $G$', or else `the affine quiver associated to the the $a$--th simple root of the group $G$'; it will be denoted as $\widehat{H}_a$,  where $a=1,2\dots, \mathrm{rank}\,G$.

\begin{figure}
\begin{center}
$$\xymatrix{\bullet \ar[rrrrrr]^{\alpha_1} &&&&&& \bullet\ar@/^0.3pc/[ddddllllll]^{\psi_1}\ar[llllldd]_{\psi_2}\\
\\
& \bullet\ar[luu]_{C_1}\ar@/^0.6pc/[rrrr]^{\alpha_3}&&&&\bullet\ar[uur]_{C_2}\ar[llllldd]^{\psi_3}\\
\\
\bullet \ar[uuuu]^{A_1} \ar[uur]^{B_1}\ar[rrrrrr]_{\alpha_3}&&&&&& \bullet\ar[uuuu]_{A_2}\ar[uul]_{B_2}}$$
$$\cw=(\alpha_1A_1-A_2\alpha_3)\psi_1+(\alpha_1C_1-C_2\alpha_2)\psi_2+(\alpha_2B_1-B_2\alpha_3)\psi_3$$
\caption{The quiver and superpotential for $\widehat{A}(2,1)\boxtimes A_2$}
\label{quoqui}
\end{center}
\end{figure}

\smallskip

In order to identify the physical models we use some invariants. The simplest invariants of a $\cn=2$ theory are the total rank $n$ of the symmetry group, equal to the number of nodes of its quiver, and the rank $f$ of its flavor symmetry group. $f$ is equal to the number of zero eigenvalues of the exchange matrix $B=S^t-S$, or equivalently, to the number of the $+1$ eigenvectors of 
the 2d monodromy $(S^{-1})^tS$  \cite{CV11}. For the $\widehat{H}\boxtimes G$ theory we have (cfr.\!\! \S.\,\ref{triangletr}) 
\begin{equation}
 (S^{-1})^tS_{\:\widehat{H}\boxtimes G}= \Phi_{\widehat{H}}\otimes \Phi_G
\end{equation}
where $\Phi_{\widehat{H}},\Phi_G$ denote the Coxeter elements of the respective Lie algebras\footnote{\ For $\widehat{A}_r$ the conjugacy class of Coxeter elements is not unique; here we mean the Coxeter class defined by the sink sequence of the $\widehat{A}(p,q)$ quiver.}. One has
\begin{gather}
 \det[\lambda-\Phi_{\widehat{H}}]= \frac{(\lambda^{p_1}-1)(\lambda^{p_2}-1)(\lambda^{p_3}-1)}{\lambda-1}
\end{gather}
where $\{p_1,p_2,p_3\}$ are the three ranks of the matter sector in table \eqref{tabletable} corresponding to $\widehat{H}$. So  $f$ is equal to the number of solutions to the equations
\begin{equation}
 \frac{\ell_i}{p_i}+\frac{k_i}{h(G)}\in\Z \qquad\qquad \begin{aligned}&i=1,2,3,\quad\ell_i=1,2,\dots, p_i-1,\\ & k_i\ \text{an exponent of }G.\end{aligned}
\end{equation}
 For instance, in the case of the model $\widehat{H}\boxtimes A_{N-1}$ this gives 
\begin{equation}\label{numflav}
 f=\gcd\{p_1,N\}+\gcd\{p_2,N\}+\gcd\{p_3,N\}-3.
\end{equation}

\subsection{Weak coupling}\label{wwweek}

We claim that the $\cn=2$ model $\widehat{H}\boxtimes G$ is SYM with gauge group $G$ coupled to some superconformal $\cn=2$ matter (which may contain further SYM sectors). The most convincing proof of this statement consists in computing the BPS mass spectrum as $g_\mathrm{YM}\rightarrow 0$ and showing that the vectors which remain light in the limit form precisely  one copy of the adjoint representation of $G$ plus, possibly, $G$--singlets. This amounts to constructing the light category $\mathscr{L}$ and checking that it has the universal structure described in \cite{cattoy}.

By standard arguments \cite{cattoy,half,nonsimply} we may choose our $S$--duality frame in such a way that the representation $X_a$, corresponding to the $a$--th simple root $W$--boson, has support in the affine quiver $\widehat{H}_a$ over the $a$--th simple root. Then, by Kac's theorem \cite{kac}, its dimension vector must be equal to the minimal imaginary roots of $\widehat{H}$
\begin{equation}
\dim X_a=\delta_a.
\end{equation}
The magnetic charges are then given by eqn.\eqref{thhffnnc}.
Since $S=S_{\widehat{H}}\otimes S_G$, this is explicitly\footnote{\ For the chain of equalities in eqn.\eqref{ccchain} see \textit{e.g} \S.10.1 and appendix A of \cite{cattoy}.}
\begin{equation}\label{ccchain}
m_a(X)=C^{-1}_{ab}\, (\delta^t S_{\widehat{H}})_i (S_{bc}+S_{cb})\dim X_{i c}\equiv \frak{d}\big(\dim X\big|_{\widehat{H}_a}\big)
\end{equation}
where $\frak{d}$ is the Dlab--Ringel defect of the (sub)quiver $\widehat{H}_a$.
 That the magnetic charges $m_a(X)$ are integrally quantized and the $W$--bosons are mutually local, $m_a(\delta_b)=0$, is a non--trivial check of our claim. The magnetic charges $m_a(\cdot)$ define the light category $\mathscr{L}$ as in \S.\,\ref{sec:light}.

By the property discussed around eqn.\eqref{tttrrxzaq}, we know that
\begin{equation}\label{mxp5}
X\in \mathscr{L}\quad \Rightarrow\quad X\big|_{\widehat{H}_a}\in \mathscr{L}(\widehat{H}).
\end{equation} 
This gives a consistency condition on the magnetic charges $m_a(\cdot)$
\begin{equation}
m_a(X)=m\big(X\big|_{\widehat{H}_a}\big)
\end{equation}
which is automatically true in view of \eqref{ccchain}. 

The category $\mathscr{L}(\widehat{H})$ is precisely the regular category $\ct$ described in eqn.\eqref{ringel}. From the list of acyclic affine quivers $\widehat{H}$ in table \eqref{tabletable} we see that each $\widehat{H}$ corresponds to a set of Argyres--Douglas matter subsector  of types $D_{p_1},\cdots D_{p_\ell}$;
then on $\mathbb{P}^1$ there are 
$\ell$ distinct points $\lambda_i$  such that
the associated category $\ct_{\lambda_i}$ is a stable tube of period $p_i$; the $\ct_\lambda$'s over all other points of  $\mathbb{P}^1$ are homogeneous tubes (period $1$).
The property \eqref{tttrrxzaq} has an important refinement. For $X\in\mathscr{L}$ one has \cite{cattoy,half,nonsimply}
\begin{equation}\label{lammmbddda}
X\big|_{\widehat{H}_a}\in \ct_\lambda(\widehat{H}) \qquad \text{the \underline{same} $\lambda$ for all }a.
\end{equation}
From \eqref{lammmbddda} it follows that the light spectrum consists of vector--multiplets in the adjoint of $G$ --- corresponding to the generic point of $\mathbb{P}^1$ --- plus the matter which resides at the special values $\lambda_i$. The family of stable representations for the $W$ boson associated to the positive root $\alpha=\sum_an_a\alpha_a$ has the following form: its restriction to $\widehat{H}_a$ is the direct sum of $n_a$ copies of the brick of dimension $\delta$ and parameter $\lambda\in\mathbb{P}^1$. The arrows connecting $\widehat{H}_a$ and $\widehat{H}_b$ vanish if oriented in one direction and are equal to the arrows in the brick of the $G$ Dynkin quiver of dimension $\alpha$ in the other direction; which of the two possible directions correspond to non--zero arrows is determined by the choice of the central charge $Z$; by comparison with pure SYM \cite{cattoy} we see the stable such states make precisely \emph{one copy} of the adjoint representation in \emph{any} weakly coupled chamber. 

The matter systems associated with two distinct special points decouple from each other as $g_\mathrm{YM}\rightarrow 0$, so, as long as we are interested in the matter theory itself rather than the full gauged model $\widehat{H}\boxtimes G$, we loose no generality in choosing $\widehat{H}$ to have just \textit{one} special point  over which we have a stable tube of period $s+1$, $s=1,2,\dots$. This corresponds to the model $\widehat{H}=\widehat{A}(s+1,1)\boxtimes G$.
Writing $D(G,s)$ for the matter theory which decouples at a special point in $\mathbb{P}^1$ such that its representations restrict to a tube of period $(s+1)$ on each affine subquiver $\widehat{H}_a$, for general $G$  the table \eqref{tabletable} gets replaced by

\begin{equation}\label{tabletable2}
\begin{tabular}{|c|c|}\hline
model & matter sector\\\hline
$\widehat{A}(p,q)\boxtimes G$\quad $p\geq q\geq1$ & $D(G,p-1)\oplus D(G,q-1)$\\\hline
$\widehat{D}_r\boxtimes G$\quad $r\geq 4$& $D(G,1)\oplus D(G,1)\oplus D(G,r-3)$\\\hline
$\widehat{E}_r\boxtimes G$\quad $r=6,7,8$& $D(G,1)\oplus D(G,2) \oplus D(G,r-4)$\\\hline 
\end{tabular}
\end{equation}

\subsection{A special model}\label{spemodel}

Let us consider the simplest $SU(3)$ gauge theory of the present class, namely $\widehat{A}(2,1)\boxtimes A_2$. Its light category is studied in great detail in appendix \ref{first appendix}. One sees that the matter sector has BPS states of spin $\leq 1/2$ in all chambers. From ref.\!\!\cite{CV11} we know that an $\cn=2$ model with this last property is either free or an Argyres--Douglas model. Given that the matter BPS spectrum has a non--trivial chamber dependence, the first possibility is ruled out. There is only one Argyres--Douglas model with a global $SU(3)$ symmetry, namely the one of type $D_4$, and we conclude that the $\widehat{A}(2,1)\boxtimes A_2$ model must be $SU(3)$ SYM coupled to Argyres--Douglas of type $D_4$. This is proven in full mathematical rigor in appendix \ref{first appendix}.
In the next section we shall give an even simpler argument for this identification.

The same result may be obtained using the approach of \cite{ACCERV2}. According to the rules of that paper, the quiver of $SU(3)$ SYM coupled to $D_4$ Argyres--Douglas is
\begin{equation}
\begin{gathered}
\xymatrix{ 1\ar@<0.5ex>[dd]\ar@<-0.5ex>[dd]&& 5\ar[ll]\ar[dr]&&&\\
&&& 6\ar[dl]&4\ar[l]\\
3\ar[rr]&&2 \ar@<0.5ex>[uu]\ar@<-0.5ex>[uu]&&&}
\end{gathered}
\end{equation}
which mutated at 4 6 4 5 2 4 gives $\widehat{A}(2,1)\boxtimes A_2$. The same argument shows that $\widehat{A}(2,2)\boxtimes A_2$ is $SU(3)$ SYM coupled to two copies of $D_4$ Argyres--Douglas, in agreement with the separation of matter systems associated to distinct $G$--tubes.

\subsection{A Lagrangian subclass}\label{laglag}

Generically the models $\widehat{H}\boxtimes G$ have no weakly coupled Lagrangian formulation. This is already true for $G=SU(2)$ \cite{CV11}. However some of them do have a Lagrangian formulation.
In particular, for the class of models $\widehat{A}(2,1)\boxtimes A_{2m-1}$ all the invariant quantities we compute agree with the ones for the quiver gauge theory
\begin{equation}\label{2mmquiv}
\xymatrix{*++[o][F-]{\;2m\;}\ar@{-}[rr] && *++[o][F-]{\ \;m\ \;}}
\end{equation}
\textit{i.e.}\! a hypermultiplet in the bifundamental $(\boldsymbol{2m},\boldsymbol{\overline{m}})$ of $SU(2m)\times SU(m)$.
\textit{E.g.}\!\! the number of nodes is $6m-3$ which is equal to the sum of $3m-2$ magnetic, $3m-2$ electric, and $1$ flavor charges for the model \eqref{2mmquiv}. Moreover, eqn.\eqref{numflav} gives $f=1$ for $\{p_1,p_2,p_3\}=\{2,1,1\}$ and $N=2m$ even. Below we shall show that also the $\beta$--function coefficient $b$ and the order of the quantum monodromy of the `matter' sector agree.
We \textit{conjecture} this identification to be correct.

Assuming the conjecture and taking the YM coupling of $SU(2m)$ to zero, the decoupled matter system $D(SU(2m),1)$ gets identified with $SU(m)$ SQCD with $N_f=2m$ flavors. This models is in facts superconformal, as predicted by our general arguments.

Let us give evidence for the conjecture. For $m=1$ it reduces to results of \cite{CV11}. For $m>1$ a valid proof requires to show that the quiver  $\widehat{A}(2,1)\boxtimes A_{2m-1}$ is mutation equivalent to
\begin{equation}
\begin{gathered}
\xymatrix{\bullet \ar[r] &\bullet\ar@<0.4ex>[dd]\ar@<-0.4ex>[dd] &\ar[l]\cdots \ar[r] &\bullet \ar@<0.4ex>[dd]\ar@<-0.4ex>[dd]&& \bullet\ar[r]\ar[ld] &\cdots &\bullet\ar[l]\\
&&&&\bullet\ar[lu]\ar[dr]\\
\bullet\ar@<0.4ex>[uu]\ar@<-0.4ex>[uu] &\bullet\ar[l]\ar[r] &\cdots &\bullet\ar[l]\ar[ur]&& \bullet\ar@<0.4ex>[uu]\ar@<-0.4ex>[uu] &\ar[l]\cdots\ar[r] &\bullet\ar@<0.4ex>[uu]\ar@<-0.4ex>[uu]}
\end{gathered}
\end{equation} 
with $2m-2$ (resp.\! $m-2$) squares on the left (resp.\! right) of the bifundamental node. 

At least for $m=2$ and $m=3$ we show explicitly that the two quivers are mutation equivalent, see appendix \ref{cecksss}.
At the level of quivers for the decoupled SCFT itself the corresponding identifications
will be shown in \S.\,\ref{sec:quiqui}. 

Another model which is Lagrangian is $\widehat{A}(3,1)\boxtimes A_2$ which corresponds to $SU(3)\times SU(2)$ SYM coupled to hypers in the representation $(\boldsymbol{3},\boldsymbol{2})\oplus (\boldsymbol{1},\boldsymbol{2})$. This can be seen in many ways, including direct mutation of the quivers, see appendix \ref{cecksss}.

\subsection{The $\beta$--function} 

The arguments of \cite{cattoy} apply to the present models; one gets for 
$\beta$--function coefficient
\begin{equation}
b(\widehat{H}\boxtimes G)= \chi(\widehat{H})\,h(G),
\end{equation}
where $\chi(\widehat{H})$ is the Euler characteristic of the domestic canonical algebra of type $\widehat{H}$, and $h(G)$ is the Coxeter number of $G$. This formula is equivalent to 
\begin{equation}
b\big(D(G,s)\big)= \frac{s}{s+1}h(G),
\end{equation}
which is between $1/4$ and $1/2$ the contribution from an adjoint hypermultiplet.
This result is consistent with the claim in the previous subsection: $b(D(SU(2m),1)=m$, which is the right value for $m$ free hypermultiplets. $b(D(SU(3),2)$ is $2$ as it should. 
In the case of the special model of \S.\,\ref{spemodel}, we get
$b(D(SU(3),1))=3/2$ which is again the right value for $D_4$ Argyres--Douglas.
The $\beta$--function may also computed by the methods of \cite{CV11}, leading to the same results.

\subsection{Strong coupling: finite BPS chambers}

For a quiver of the form $Q\boxtimes G$, where $Q$ is acyclic and $G$ is Dynkin, a \emph{finite} BPS chamber  containing only hypermultiplets with charge vectors
\begin{equation}
e_a\otimes \alpha \in \Gamma_Q\otimes \Gamma_G, \qquad \alpha\in \Delta^+(G),
\end{equation}  
that is, a copy of the positive roots of $G$ for each node of $Q$. This result is well--known for the $G\,\square\,G^\prime$ models \cite{CNV} and may be proven for all pairs of acyclic quivers.
We get a finite chamber with
$$\#\{\text{hypermultiplets}\}=\frac{1}{2}\,\mathrm{rank}\,\widehat{H}\:\mathrm{rank}\,G\:h(G).$$

\section{The SCFT models $D(G,s)$}

\subsection{Quivers and superpotentials}\label{sec:quiqui}

We have natural candidate quivers for $D(G,s)$, namely the 
`$\widehat{A}(s+1,0)\boxtimes G$' ones. For each node $a\in G$ `$\widehat{A}(s+1,0)\boxtimes G$' has a full subquiver which is an oriented simple $(s+1)$--cycle. Two such cycles are connected iff the corresponding nodes are connected in the Dynkin quiver $G$; they are connected by arrows of the form $e_i\otimes \eta$ as well as by the arrows $\psi$ implementing the commutativity relations (cfr.\! \!\S.\,\ref{triangletr}). We stress that the resulting quiver is not necessarily $2$--acyclic (see \cite{ACCERV2} for a discussion).

The superpotential has the  form
\begin{equation}\label{superWsuper}
\cw= \sum \{(s+1)\text{--cycles}\}+\sum \psi\text{(commutators)}+\cdots
\end{equation}
where the ellipsis stand for higher order terms that we cannot rule out, but expect not to be present or relevant. 

\subsubsection{Flavor group}

An important check on the proposed quiver is that the theory it describes has a flavor group $F\supseteq G$. Let us start by computing the rank of the flavor group $F$. The exchange matrix is
\begin{equation}
B=(\boldsymbol{1}-P)\otimes S_G^t-(\boldsymbol{1}-P^{-1})\otimes S_G,
\end{equation}
where $P$ is the cyclic permutation $(s+1)\times(s+1)$ matrix. All vector of the form $1\otimes v$ are zero eigenvectors of $B$. They correspond to the $\mathrm{rank}\,G$ charges associated to the Cartan of $G$; in addition we have the flavor symmetries already present in the $\widehat{A}(s+1,1)\boxtimes G$ theory
\begin{equation}\label{tttyyuui}
\mathrm{rank}\,F=\mathrm{rank}\, G+\sum_{d\mid  (s+1)\atop d\in I(G)} \varphi(d),\end{equation}  
where $\varphi(d)$ is the Euler totient function and $I(G)$ is the set 
\begin{align*}
&I(A_r)=\{d\:\colon\: d\mid (r+1),\ d>1\}\qquad I(D_r)=\{2\}\cup\{ d\:\colon\; d\mid 2(r-1),\ d\neq 1,r-1\}\\
&I(E_6)=\{3,12\},\qquad I(E_7)=\{2,18\},\qquad I(E_8)=\{30\}.
\end{align*}
For instance, consider the two models $D(SU(4),1)$ and $D(SU(3),2)$ both corresponding to $SU(2)$ SQCD with $N_f=4$; one gets
\begin{equation}
\mathrm{rank}\,F=\mathrm{rank}\,SU(4)+\varphi(2)=\mathrm{rank}\,SU(3)+\varphi(3)=4\equiv \mathrm{rank}\,SO(8).
\end{equation}
Next we argue that we have the group $G$ and not just its Cartan subgroup. Given a stable representation $X$ of our quiver, we may extend it to a stable representation of $\widehat{A}(s+1,1)\boxtimes G$ which belongs to the $G$--tube. The stable representations of $\widehat{A}(s+1,1)\boxtimes G$ are organized in representations of the gauge group $G$; the representations in the `orbit' of $X$ belong to the same $G$--tube and may be identified with stable representations of the matter quiver. 

A consequence of eqn.\eqref{tttyyuui} is that the models $D(E_8,s)$ with $30\nmid (s+1)$ and $D(E_6,s)$ with $3\nmid (s+1)$ have \emph{exactly} flavor group $E_8$ and, respectively, $E_6$. On the other hand, for $s$ odd $D(E_7,s)$ has always  a symmetry strictly larger than $E_7$.

\subsubsection{Examples and checks}

We check the above assertions in a number of examples. The first examples are the $D(SU(2),s)$ models, that is, Argyres--Douglas of type $D_{s+1}$; the equivalence with the $\widehat{A}(s+1,0)\boxtimes A_1$ quiver (with $\cw$ as in \eqref{superWsuper}) is shown in \cite{CV11}. 
\smallskip

Next let us consider $D(SU(3),1)$. According to \S.\,\ref{spemodel} it must be Argyres--Douglas of type $D_4$. The
$\widehat{A}(2,0)\boxtimes A_2$ quiver is
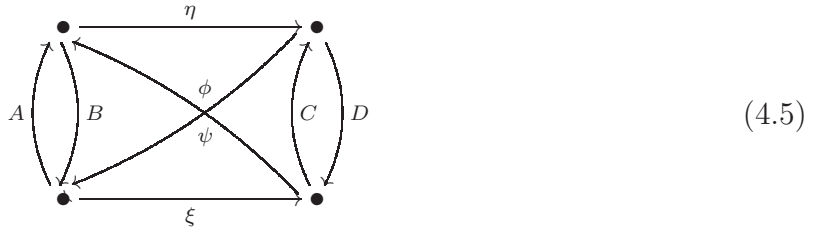
\begin{equation}\label{firstexampleeee}
 \begin{gathered}
  \xymatrix{\bullet\ar@<-0.8ex>@/^0.8pc/[dd]^B \ar[rrr]^\eta &&& \bullet \ar@/^0.8pc/[dd]^D\ar@<-0.8ex>@/^0.6pc/[ddlll]^\psi\\
\\
\bullet\ar\ar@<0.4ex>@/^0.8pc/[uu]^A \ar[rrr]_\xi &&& \bullet\ar@<0.8ex>@/_0.6pc/[uulll]_\phi \ar@/^0.8pc/[uu]_C}
 \end{gathered}
\end{equation}
\begin{equation}
 \cw= AB+CD+ \psi(\xi B-D\eta)+\phi(\eta A-C\xi).
\end{equation}
Eliminating the massive `Higgs fields' $A,B,C,D$ trough their equations of motion we get
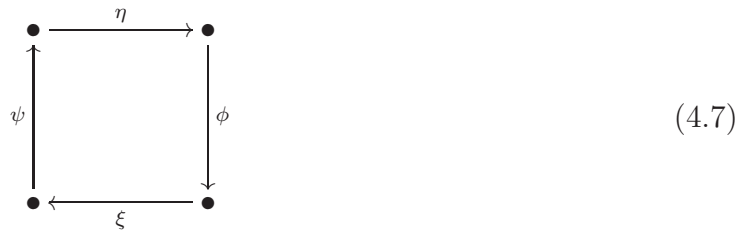
\begin{equation}
 \begin{gathered}
 \xymatrix{\bullet\ar[rr]^\eta && \bullet\ar[dd]^\phi\\
\\
\bullet\ar[uu]^\psi &&\bullet\ar[ll]^\xi} 
 \end{gathered}
\end{equation}
with superpotential $\cw=-2\psi\xi\phi\eta$, which is precisely the $D_4\equiv A_2\,\square\,A_2$ model, as expected.
In the same vein, we may check the models $D(SU(4),1)$ and $D(SU(6),1)$ which we know to correspond, respectively, to 
$SU(2)$ SQCD with $N_f=4$ and $SU(3)$ SQCD with $N_f=6$. As in the example \eqref{firstexampleeee}, for all $D(G,1)$ models, the quiver $\widehat{A}(2,0)\boxtimes G$ has massive $2$--cycles  which may be integrated away. For $D(SU(4),1)$ we remain with the quiver
\begin{equation}
 \xymatrix{\circ\ar[rr] && \bullet\ar[dll]\ar[drr] && \circ\ar[ll]\\
\bullet\ar[rr] && \bullet\ar[ull]\ar[urr] && \bullet\ar[ll] }
\end{equation}
Mutation at the two white nodes $\circ$ transforms this quiver in the standard one for $SU(2)$ SQCD with $N_f=4$.
The quiver of the $D(SU(6),1)$ model is
\begin{equation}
\xymatrix{2\ar[r] &4\ar[r]\ar[dl]& 6\ar[r]\ar[dl] & 8\ar[r] \ar[dl]&10\ar[dl]\\
1\ar[r] &3\ar[r]\ar[ul]& 5\ar[r]\ar[ul] & 7\ar[r]\ar[ul] &9\ar[ul] }
\end{equation}
whose mutation at nodes 7 2 8 7 9 4 8 10 7 3 8 10 6 8 6 8 produces
\begin{equation}
\begin{gathered}
\xymatrix{
&&&4\ar@<0.5ex>[dd]\ar@<-0.5ex>[dd]&&\ar[ll]8\ar[dr]\ar[drr]\ar[drrr]&&&\\
3\ar[urrr]&2\ar[urr]&1\ar[ur]&&&&6\ar[dl]&7\ar[dll]&10\ar[dlll]\\
&&&5\ar[ul]\ar[ull]\ar[ulll]\ar[rr]&&9\ar@<0.5ex>[uu]\ar@<-0.5ex>[uu]&&&\\
}
\end{gathered}
\end{equation}
which is the quiver of $SU(3)$ $N_f=6$. The quiver for $D(SU(3),2)$ is mutation--finite, and by classification it is easily identified with the one for $SU(2)$ SQCD with $N_f=4$, in agreement with our previous findings.
These examples provide strong evidence that the obvious candidate quiver (with superpotential) is indeed the correct one.

\subsection{Order of the quantum monodromy, dimension of chiral fields}\label{vvvmmt451}
Since the theory $D(G,s)$ is $\cn=2$ superconformal, its quantum monodromy $M(q)$ has finite order $r$ \cite{CNV}, and all chiral primary operators have dimensions in $\mathbb{N}/r$. The order $r$ is a nice invariant which is quite useful to distinguish SCFT models.

\smallskip
 
Repeating the scaling arguments at the end of \S.\,\ref{AFEucli}, we see that the matter theory $D(G,s)$, at the formal level, is engineered by the local Calabi--Yau geometry
\begin{equation}\label{surfhyper}
W\equiv e^{(s+1)z}+W_G(x_1,x_2,x_3)=0
\end{equation}
endowed with the standard holomorphic $3$--form
\begin{equation}
\Omega= P.R.\,\frac{dz \wedge dx_1\wedge dx_2\wedge dx_3}{W}
\end{equation}
($P.R.$ stands for `Poincar\'e Residue'). At a conformal point $W_G(x_i)$ is quasi--homogeneous, $W_G(\lambda^{q_i}x_i)=\lambda\,W_G(x_i)$ for all $\lambda\in \C$. Thus 
\begin{equation}
x_i\rightarrow e^{i \alpha q_i}\,x_i,\qquad z\rightarrow z+\alpha/(s+1), 
\end{equation}
is a holomorphic symmetry of the hypersurface \eqref{surfhyper} under which
\begin{equation}
\Omega \rightarrow \exp\big(i\alpha(q_1+q_2+q_3-1)\big)\Omega,
\end{equation}
so that the dimension of $x_i$ is $q_i/(\sum_jq_j-1)\equiv q_i\, h(G)\in \Z$ while that of $e^z$ is $h(G)/(s+1)$.
\smallskip

The order of the quantum monodromy $M(q)$ is then
\begin{equation}
\text{order }M(q)\equiv r=\frac{s+1}{\gcd\{s+1, h(G)\}}.
\end{equation}

Let us check that this formula reproduces the right results for the special models. For $D(SU(2),s)$ we get 
\begin{equation}r= \begin{cases}s+1 & s+1\ \text{odd}\\
(s+1)/2 & s+1\ \text{even}\end{cases}=\begin{cases}(h(D_{s+1})+2)/\gcd\{h(D_{s+1}),2\}\\
(h(D_{s+1})+2)/[2\gcd\{h(D_{s+1}),2\}] \end{cases}\end{equation}
the \textsc{rhs} is the monodromy order for the Argyres--Douglas model of type $D_{s+1}$ \cite{CNV}, in agreement with our identification of this model with the $D(SU(2),s)$ one. Likewise, for $D(SU(3),1)$
$$r=2\equiv \frac{h(A_2)+h(A_2)}{\gcd\{h(A_2),h(A_2)\}},$$
consistent with the fact that $D(SU(3),1)\sim A_2\,\square\,A_2$, while for $D(SU(4),1)$ we get $r=1$ as it should be for a model having a Lagrangian formulation. For $D(SU(2m),1)$ we also get $r=1$ as expected if our conjecture holds.

\subsection{Lagrangian models?}

We may ask which of the models $D(G,s)$ may possibly have a Lagrangian description. A necessary condition for a SCFT $D(G,s)$ to have a Lagrangian formulation is
\begin{equation}
 r(D(G,s))=1, \qquad b(D(G,s))\in\mathbb{N}.
\end{equation}
Each of the two conditions is satisfied iff $(s+1)\mid h(G)$. Note that $r=1\ \Leftrightarrow\ b\in\mathbb{N}$. Morever, for $G=SU(N)$, $r=1$ also implies $f\geq 1$, as required in a $\cn=2$ Lagrangian gauge theory with matter in a  representation $R\oplus\overline{R}$ ($R$ complex) of the gauge group $K$. 

Whenever $(s+1)\nmid h(G)$ the model $D(G,s)$ is necessarily intrinsically strongly coupled.

\section{$\widehat{H}\boxtimes \widehat{G}$ models}

In the previous constructions we  used that Type IIB superstring on a 3--CY hypersurface with $\hat c<2$ produces a \emph{bona fide} 4D $\cn=2$ QFT. It is believed that this remains true if the upper bound is saturated, $\hat c=2$.
In this case we can consider models of the form $\widehat{H}\boxtimes \widehat{G}$ with $\widehat{H},\widehat{G}$ two (acyclic) affine quivers, which automatically has $\hat c=2$ \cite{CV92}. The corresponding QFT are expected to be asymptotically--free gauge theories.

An analysis of these more general models is beyond the scope of this letter. We plan to return to them in a separate publication. The theory has a large flavor group $F$; if $\{p_1,p_2,p_3\}$ (resp.\! $\{q_1,q_2,q_3\}$) are the periods corresponding to $\widehat{H}$ (resp.\! $\widehat{G}$) in table \eqref{tabletable}, one has
\begin{equation}
\mathrm{rank}\,F= \sum_{1\leq i,j\leq 3}\Big(\gcd\{p_i,q_j\}-1\Big).
\end{equation}

 In order to extract a SCF `matter' sector we may think of decoupling the SYM sectors.
Starting with the prototypical such AF theory, $\widehat{A}(s+1,1)\boxtimes\widehat{A}(t+1,1)$, which corresponds geometrically to the hypersurface
\begin{equation}
 W\equiv \Lambda^b\, e^{(s+1)z}+\Lambda^b\,e^{-z}+\widetilde{\Lambda}^{b^\prime}\,e^{(t+1)y}+\widetilde{\Lambda}^{b^\prime}\,e^{-y}+x_1^2+x_2^2=0,
\end{equation}
 one would expect  to end up with a putative SCF matter theory described by the would--be `$\widehat{A}(s+1,0)\boxtimes \widehat{A}(t+1,0)$' quiver.
However, the counting of nodes now works differently. The number of magnetic charges which disappear in the double weak--coupling limit is
$$(s+2)+(t+2)=s+t+4,$$
while the difference in the number of nodes between the quivers $\widehat{A}(s+1,1)\boxtimes\widehat{A}(t+1,1)$ and
 $\widehat{A}(s+1,0)\boxtimes \widehat{A}(t+1,0)$ is
$$(s+2)(t+2)-(s+1)(t+1)=s+t+3,$$
so we have a mismatch by one node. There is an obvious way of decreasing by one the rank of a quiver: take the (quiver of the) category controlled by the function $\lambda(X)=\dim X_\ast$, where $\ast$ is some node of the quiver \cite{cattoy}. In the case of the quiver $\widehat{A}(s+1,0)\boxtimes \widehat{A}(t+1,0)$ all nodes are equivalent  so all choices lead to the same quiver.
The resulting theory is expected to have (at least) a symmetry ($N=s+1,M=t+1$)
\begin{equation}
SU(N)\times SU(M)\times \text{(a group of rank $\gcd\{N,M\}-1$)}.
\end{equation}
For $M=N$ the quivers
 coincide with the ones described in section 6 of ref.\!\!\cite{ACCERV2} for the Gaiotto theory $\ct_N$ corresponding to 6D $A_{N-1}$ (2,0) compactified on a sphere with three maximal punctures which has flavor symmetry (at least) $SU(N)^3$.  
 For the non--diagonal case, the quivers may be seen as arising from a circle compactification of a 5d web \cite{ben}.
\medskip

\section*{Acknowledgements}

We have greatly benefited of discussions with
Murad Alim,  Clay C\'ordova, Sam Espahbodi, Berhard Keller,
Ashwin Rastogi and Cumrun Vafa.  We thank them all.
S.C. thanks the Department of Physics of Harvard University, where this works was finished, for hospitality.

\appendix

\section{Detailed study of the light category of $\widehat{A}(2,1)\boxtimes A_2$}\label{first appendix}

The quiver and superpotential for this model are presented in figure \ref{quoqui}.
If we are interested in the subcategory $\mathcal{L}$, by eqn.\eqref{mxp5} we can take $A_1,A_2$ to be isomorphisms and identify nodes pairwise trough them. Then the fields $\psi_1$ and $\alpha_1-\alpha_2$ get massive and may be integrated away. We remain with the quiver and superpotential
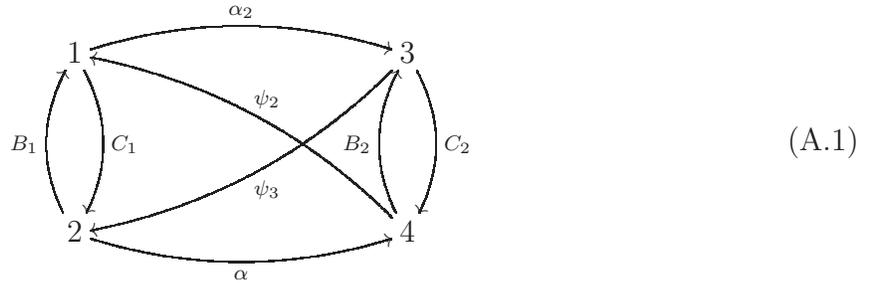
\begin{equation}\label{lightquivvv}\begin{gathered}\xymatrix{1\ar@/^0.9pc/[dd]^{C_1} \ar@/^0.9pc/[rrrr]^{\alpha_2}&&&& 3\ar@/^0.9pc/[dd]^{C_2}\ar@/^1pc/[ddllll]^{\psi_3}\\
\\
2\ar@/^0.9pc/[uu]^{B_1}\ar@/_0.9pc/[rrrr]_\alpha &&&& 4\ar@/^0.9pc/[uu]^{B_2}\ar@/_1pc/[uullll]_{\psi_2}}\end{gathered}\end{equation}
$$\cw_\mathrm{eff}=\psi_2(\alpha C_1-C_2\alpha_2)+\psi_3(\alpha_2B_1-B_2\alpha).$$
The following map is an element of $\mathrm{End}\,X$
\begin{equation}
(X_1,X_2,X_3,X_4)\mapsto (B_1C_1X_1,C_1B_1X_2, B_2C_2X_3,C_2B_2X_4)
\end{equation}
hence a complex number $\lambda$ if $X$ is a brick. For $\lambda\neq 0$, $B_1,B_2,C_1,C_2$ are isomorphisms, which  identify the nodes in pairs. The arrows $\alpha-\alpha_2$ and $\psi_2-\psi_3$ also get massive and may be integrated away, reducing to representations of the preprojective algebra $\cp(A_2)$, \textit{i.e.}\, to the homogeneous $SU(3)$--tube \cite{cattoy}. At $\lambda=0$ we isolate the non--homogeneous $SU(3)$--tube containing the matter. It corresponds to the representations of the quiver \eqref{lightquivvv} bounded by the relations
\begin{gather}\label{hhhha1}
B_1C_1=C_1B_1=B_2C_2=C_2B_2=\psi_2\alpha=\alpha_2\psi_2=\psi_3\alpha_2=\alpha\psi_3=0\\
C_1\psi_2-\psi_3B_2=\psi_2C_2-B_1\psi_3=\alpha C_1-C_2\alpha_2=\alpha_2B_1-B_2\alpha=0.\label{hhhha2}
\end{gather}

\textbf{Theorem.} \textit{The bricks $X$ of the quiver \eqref{lightquivvv} bounded by the relations \eqref{hhhha1}\eqref{hhhha2} are isolated (no moduli). They satisfy
\begin{equation}
\dim X\leq (1,1,1,1)
\end{equation}
with equality only for modules in the projective closure of the families of representations of the gauge vectors. The dimension vectors of
bricks coincide with  those for
$\C \widehat{A}(4,0)/(\partial[\text{4--cycle}])$.}\smallskip

\textbf{Proof.} By virtue of the relations in the first line, eqn.\eqref{hhhha1}, our algebra $\mathscr{A}$ is a \emph{string} algebra. In view of the
Butler--Ringel theorem \cite{butring}, the bricks of $\mathscr{A}$ are isolated iff there is no band which is a brick.
In any legitimate string, arrows (direct or inverse) labelled by latin and greek letters alternate.
We observe that a sequence of three arrows (direct of inverse) of the form (latin)(greek)(latin) is not legitimate unless the greek arrow points in the opposite direction with respect to the latin ones [same with (latin) $\leftrightarrow$ (greek)]. Indeed by  \eqref{hhhha2}
\begin{align*}
&\xrightarrow{C_1}\xrightarrow{\alpha}\xrightarrow{B_2}\ =\ \xrightarrow{C_1}\xrightarrow{B_1}\xrightarrow{\alpha_2}
&&\xrightarrow{C_1}\xrightarrow{\alpha}\xleftarrow{C_2}\ =\ \xrightarrow{\alpha_2}\xrightarrow{C_2}\xleftarrow{C_2}
&&\xrightarrow{C_1}\xleftarrow{\psi_3}\xleftarrow{B_2}\ =\ \xrightarrow{C_1}\xleftarrow{C_1}\xleftarrow{\psi_2}
\end{align*}
and the \textsc{rhs} are illegitimate strings. Thus, for all indecomposables of total dimension $\sum_i \dim X_i\geq 4$, the arrows in the string/band should alternate both in alphabets (latin vs.\!\! greek) and orientation (direct vs.\!\! inverse). Then, given an arrow in the string, the full sequence of its successors is uniquely determined. There are no bands with $\dim X_1=0$; if $\dim X_1\neq 0$ we may cyclically rearrange the band in such a way that the first node is $1$ and the first arrow is latin. If it is $C_1$, the unique continuation of the string is
\begin{equation}\label{jhjhklf}
1\xrightarrow{C_1} 2 \xleftarrow{\psi_3} 3 \xrightarrow{C_2} 4 \xleftarrow{\,\alpha\,} 2\xrightarrow{B_1} 1,
\end{equation}  
while, if the first arrow is $B_1$, it is this string segment read from the right. We cannot close \eqref{jhjhklf} to make a band since $C_1B_1=0$.
The string/band may be continued (either ways)
\begin{equation}\label{mmssdff}
\cdots \xleftarrow{\alpha_2} 1\xrightarrow{C_1} 2 \xleftarrow{\psi_3} 3 \xrightarrow{C_2} 4 \xleftarrow{\,\alpha\,} 2\xrightarrow{B_1} 1\xleftarrow{\psi_2} 4 \xrightarrow{B_2} 3 \xleftarrow{\alpha_2} 1 \xrightarrow {C_1} \cdots,
\end{equation}  
and this structure repeats periodically; all legitimate strings are substrings of a $k$--fold iteration of the period.
Let $v_i$ be the basis elements of $X_1$ numbered according to their order along the string; from \eqref{mmssdff} we see that
$v_1\mapsto v_1+v_2$,  $v_i\mapsto v_i$ for $i\geq 2$,
is a non--trivial endomorphism, so the corresponding string/band module $X$ is not a brick.  $X$ may be a brick only if $\dim X_1\leq 1$; the nodes being all equivalent, $\dim X_i\leq 1$ for all $i$. Now it is elementary to show that the matter category has a quiver and superpotential equal to those of $D_4$ \cite{cattoy}.\hfill $\square$

\section{Checks of \S 3.3.}\label{cecksss}
The quiver of $A(2,1)\boxtimes A_3$ is:
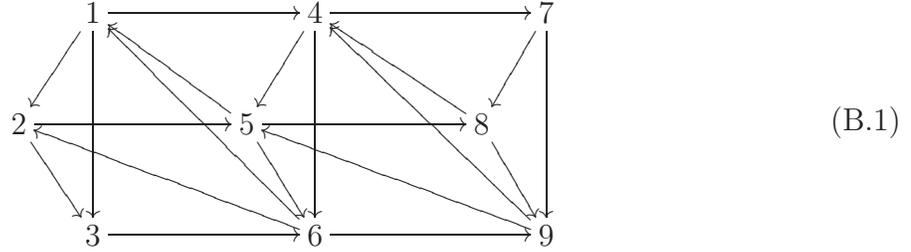
\begin{equation}
\begin{gathered}
\begin{xy} 0;<0.7pt,0pt>:<0pt,-0.7pt>:: 
(40,0) *+{1} ="0",
(0,60) *+{2} ="1",
(40,120) *+{3} ="2",
(160,0) *+{4} ="3",
(123,60) *+{5} ="4",
(160,120) *+{6} ="5",
(285,0) *+{7} ="6",
(250,60) *+{8} ="7",
(285,120) *+{9} ="8",
"0", {\ar"1"},
"0", {\ar"2"},
"0", {\ar"3"},
"4", {\ar"0"},
"5", {\ar"0"},
"1", {\ar"2"},
"1", {\ar"4"},
"5", {\ar"1"},
"2", {\ar"5"},
"3", {\ar"4"},
"3", {\ar"5"},
"3", {\ar"6"},
"7", {\ar"3"},
"8", {\ar"3"},
"4", {\ar"5"},
"4", {\ar"7"},
"8", {\ar"4"},
"5", {\ar"8"},
"6", {\ar"7"},
"6", {\ar"8"},
"7", {\ar"8"},
\end{xy}
\end{gathered}
\end{equation}
Mutating at the nodes 7 4 8 2 5 9 4 6 9 6 7 6 4 8 we obtain:
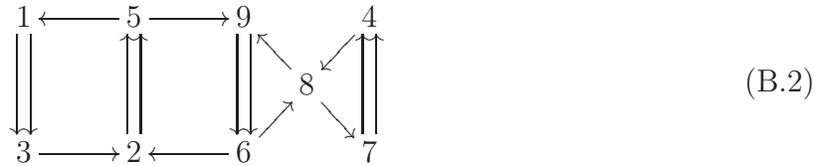
\begin{equation}
\begin{gathered}
\xymatrix@R=0.99pc@C=0.99pc{
1\ar@<0.5ex>[dd]\ar@<-0.5ex>[dd]&&5\ar[rr]\ar[ll]&&9\ar@<0.5ex>[dd]\ar@<-0.5ex>[dd]&&4\ar[dl]\\
&&&&&8\ar[ul]\ar[dr]&\\
3\ar[rr]&&2\ar@<0.5ex>[uu]\ar@<-0.5ex>[uu]&&6\ar[ll]\ar[ur]&&7\ar@<0.5ex>[uu]\ar@<-0.5ex>[uu]
}
\end{gathered}
\end{equation}
The quiver of $A(2,1)\boxtimes A_5$ is
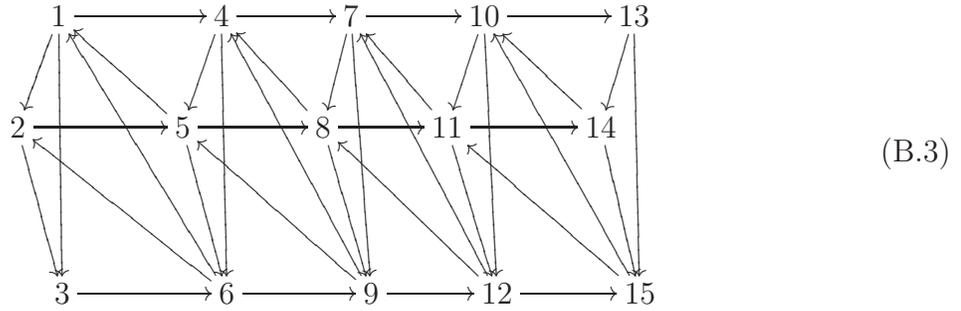
\begin{equation}
\begin{gathered}
\begin{xy} 0;<0.7pt,0pt>:<0pt,-0.7pt>:: 
(22,0) *+{1} ="0",
(0,60) *+{2} ="1",
(24,150) *+{3} ="2",
(110,0) *+{4} ="3",
(89,60) *+{5} ="4",
(113,150) *+{6} ="5",
(180,0) *+{7} ="6",
(165,60) *+{8} ="7",
(191,150) *+{9} ="8",
(252,0) *+{10} ="9",
(232,60) *+{11} ="10",
(259,150) *+{12} ="11",
(333,0) *+{13} ="12",
(315,60) *+{14} ="13",
(336,150) *+{15} ="14",
"0", {\ar"1"},
"0", {\ar"2"},
"0", {\ar"3"},
"4", {\ar"0"},
"5", {\ar"0"},
"1", {\ar"2"},
"1", {\ar"4"},
"5", {\ar"1"},
"2", {\ar"5"},
"3", {\ar"4"},
"3", {\ar"5"},
"3", {\ar"6"},
"7", {\ar"3"},
"8", {\ar"3"},
"4", {\ar"5"},
"4", {\ar"7"},
"8", {\ar"4"},
"5", {\ar"8"},
"6", {\ar"7"},
"6", {\ar"8"},
"6", {\ar"9"},
"10", {\ar"6"},
"11", {\ar"6"},
"7", {\ar"8"},
"7", {\ar"10"},
"11", {\ar"7"},
"8", {\ar"11"},
"9", {\ar"10"},
"9", {\ar"11"},
"9", {\ar"12"},
"13", {\ar"9"},
"14", {\ar"9"},
"10", {\ar"11"},
"10", {\ar"13"},
"14", {\ar"10"},
"11", {\ar"14"},
"12", {\ar"13"},
"12", {\ar"14"},
"13", {\ar"14"},
\end{xy}
\end{gathered}
\end{equation}
Mutation at  13 10 14 7 11 15 4 8 12 13 2 5 6 4 9 7 9 6 8 7 9 12 7 5 3 10 12 9 10 6 12 10 7 12 7 11 7 12 6 9 13 15 14 11 13 12 8 5 3 15 12 8 5 3 15 11 9 14 6 14 9 6 14 6 14 9 7 14 11 gives:
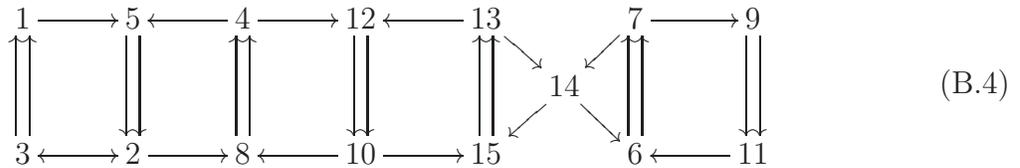
\begin{equation}
\begin{gathered}
\xymatrix@R=0.99pc@C=0.99pc{1\ar[rr]&&5\ar@<0.5ex>[dd]\ar@<-0.5ex>[dd]&&4\ar[ll]\ar[rr]&&12\ar@<0.5ex>[dd]\ar@<-0.5ex>[dd]&&13\ar[dr]\ar[ll]&&7\ar[dl]\ar[rr]&&9\ar@<0.5ex>[dd]\ar@<-0.5ex>[dd]\\
&&&&&&&&&14\ar[dl]\ar[dr]&&&\\
3\ar@<0.5ex>[uu]\ar@<-0.5ex>[uu]\ar[rr]&&2\ar[ll]\ar[rr]&&8\ar@<0.5ex>[uu]\ar@<-0.5ex>[uu]&&10\ar[ll]\ar[rr]&&15\ar@<0.5ex>[uu]\ar@<-0.5ex>[uu]&&6\ar@<0.5ex>[uu]\ar@<-0.5ex>[uu]&&11\ar[ll]\\}
\end{gathered}
\end{equation}

The quiver of $SU(3)\times SU(2)$ coupled to $(\mathbf{3,2})\oplus (\mathbf{1,2})$ is
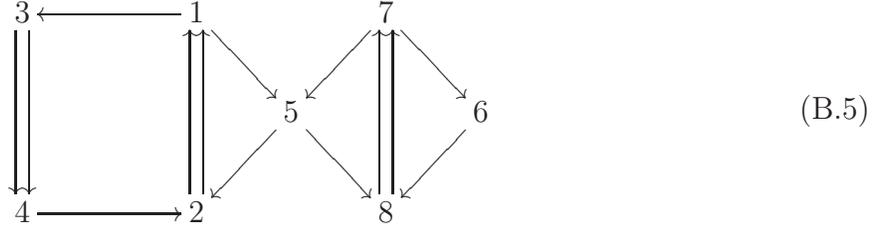
\begin{equation}
\begin{gathered}
\xymatrix{ 3\ar@<0.5ex>[dd]\ar@<-0.5ex>[dd]&& 1\ar[ll]\ar[dr]&&7\ar[dl]\ar[dr]&\\
&&& 5\ar[dl]\ar[dr]&&6\ar[dl]\\
4\ar[rr]&&2 \ar@<0.5ex>[uu]\ar@<-0.5ex>[uu]&&8\ar@<0.5ex>[uu]\ar@<-0.5ex>[uu]&}
\end{gathered}
\end{equation}
 by the sequence of mutations 5 8 3 2 4 8 7 1 6 8 4 5 2 it becomes the quiver $A(3,1) \boxtimes A_2$. 

 \end{document}